\documentclass[graybox, oneside, openany]{svmult}

\usepackage{type1cm}        % activate if the above 3 fonts are
\usepackage{makeidx}         % allows index generation
\usepackage{graphicx}        % standard LaTeX graphics tool
\usepackage{multicol}        % used for the two-column index
\usepackage[bottom]{footmisc}% places footnotes at page bottom

\usepackage{newtxtext}       % 
\usepackage{newtxmath}       % selects Times Roman as basic font

\usepackage[square,sort,comma,numbers]{natbib}
\usepackage{titlesec, titlecaps}
\titleformat{\chapter}[display]
{\normalfont\Large\bfseries}{\chaptertitlename~\thechapter}{20pt}{\Large}
\titlespacing*{\chapter}{0pt}{30pt}{20pt}
\renewcommand{\cleardoublepage}{}
\renewcommand{\clearpage}{}
\makeindex             % used for the subject index
\definecolor{mic}{rgb}{0.9, 0.0, 0.6}

\definecolor{maxb}{rgb}{0, 0, 153}

\begin{document}
\title*{Periodic Astrometric Signal Recovery through Convolutional Autoencoders}
\titlerunning{Astrometric Signal Recovery with Deep Learning}

\author{Michele Delli Veneri, Louis Desdoigts, Morgan A. Schmitz, Alberto Krone-Martins, Emille E. O. Ishida, Peter Tuthill, Rafael S. de Souza, Richard Scalzo, Massimo Brescia, Giuseppe Longo, Antonio Picariello}
\authorrunning{Delli Veneri et al. 2020}

\institute{M.Delli Veneri \at University of Naples Federico II, DIETI \email{michele.delliveneri@unina.it} \and L. Desdoigts \at School of Physics, The University of Sydney, NSW 2006, Australia \email{louis.desdoigts@sydney.edu.au} \and M. A. Schmitz \at Department of Astrophysical Sciences, Princeton University, 4 Ivy Ln., Princeton, NJ08544, USA \email{morgan.schmitz@astro.princeton.edu} \and A. Krone-Martins \at Donald Bren School of Information and Computer Sciences, University of California, Irvine, CA 92697, USA \at CENTRA/SIM, Faculdade de Ci\^encias, Universidade de Lisboa, Ed. C8, Campo Grande, 1749-016, Lisboa, Portugal \email{algol@uci.edu}, \and P. Tuthill \at School of Physics, The University of Sydney, NSW 2006, Australia \email{peter.tuthill@sydney.edu.au} \and E. E. O. Ishida \at Universit\'e Clermont Auvergne, CNRS/IN2P3, LPC, F-63000 Clermont-Ferrand, France \email{emille.ishida@clermont.in2p3.fr} \and R. S. de Souza \at Key Laboratory for Research in Galaxies and Cosmology, Shanghai Astronomical Observatory, Chinese Academy of Sciences, 80 Nandan Road, Shanghai 200030, China \email{drsouza@shao.ac.cn} \and R. Scalzo \at Centre for Translational Data Science, University of Sydney, Darlington NSW 2008 \email{richard.scalzo@sydney.edu.au} \and G. Longo \at University of Naples Federico II, Department of Physics E. Pancini \email{giuseppe.longo@unina.it} \and A. Picariello \at University of Naples Federico II, DIETI \email{antonio.picariello@unina.it} \and M. Brescia \at INAF - Astronomical Observatory of Capodimonte \email{massimo.brescia@inaf.it}}

\maketitle

\emph{Preprint version of the manuscript to appear in the Volume ``Intelligent Astrophysics''
of the series ``Emergence, Complexity and Computation'', Book eds. I. Zelinka, D.
Baron, M. Brescia, Springer Nature Switzerland, ISSN: 2194-7287\\}
\abstract*{
Astrometric detection involves a precise measurement of stellar positions, and is widely regarded as the leading concept presently ready to find earth-mass planets in temperate orbits around nearby sun-like stars
    The TOLIMAN space telescope \citep{toliman_proc} is a low-cost, agile mission concept dedicated to narrow-angle astrometric monitoring of bright binary stars. In particular the mission will be optimised to search for habitable-zone planets around  $\alpha$~Centauri~AB. If the separation between these two stars can be monitored with sufficient precision, tiny perturbations due to the gravitational tug from an unseen planet can be witnessed and, given the configuration of the optical system, the scale of the shifts in the image plane are about one millionth of a pixel. Image registration at this level of precision has never been demonstrated (to our knowledge) in any setting within science. In this paper we demonstrate that a Deep Convolutional Auto-Encoder is able to retrieve such a signal from simplified simulations of the TOLIMAN data and we present the full experimental pipeline to recreate out experiments from the simulations to the signal analysis. In future works, all the more realistic sources of noise and systematic effects present in the real-world system will be injected into the simulations.
}

\abstract{
        Astrometric detection involves a precise measurement of stellar positions, and is widely regarded as the leading concept presently ready to find earth-mass planets in temperate orbits around nearby sun-like stars.
    The TOLIMAN space telescope \citep{toliman_proc} is a low-cost, agile mission concept dedicated to narrow-angle astrometric monitoring of bright binary stars. In particular the mission will be optimised to search for habitable-zone planets around  $\alpha$~Centauri~AB. If the separation between these two stars can be monitored with sufficient precision, tiny perturbations due to the gravitational tug from an unseen planet can be witnessed and, given the configuration of the optical system, the scale of the shifts in the image plane are about one millionth of a pixel. Image registration at this level of precision has never been demonstrated (to our knowledge) in any setting within science. In this paper we demonstrate that a Deep Convolutional Auto-Encoder is able to retrieve such a signal from simplified simulations of the TOLIMAN data and we present the full experimental pipeline to recreate out experiments from the simulations to the signal analysis. In future works, all the more realistic sources of noise and systematic effects present in the real-world system will be injected into the simulations.
}
%\tableofcontents
\section{Introduction}
\label{sec:introduction}

Astronomy seeks to answer our deepest questions. Where did it all begin and how is it going to end? Are we alone in the Universe? Is there life beyond our biosphere -- or conversely is Earth and our planetary system in some way unique? Such inquiries have given rise to the fields of astrobiology and exoplanetary research.

Despite our long term commitment to explore these questions, the development of instruments capable of detecting planets around distant stars has proven to be one of the most challenging astronomical quests \citep{Lee_2018}. The first exoplanet orbiting a Sun-like star was detected through small deviations caused in radial velocity measurements of its host \citep[][this work  was subsequently awarded the 2019 Nobel Prize in Physics]{Mayor1995}.  A little more than twenty years later, there are more than 4000 confirmed exoplanets\footnote{\url{http://exoplanet.eu/catalog/}}.
The celestial garden is therefore a fertile ground for discovery, and the synergy between new astronomical missions and modern statistical learning techniques promises an exceptionally bright future for this rapidly expanding field. 
Discovery and characterisation of exoplanets is particularly suited to combinations of approaches that are able to push the boundaries in both the acquisition of exceptionally clean, low-noise data, as well as the ability to sift large volumes of observations in order to extract subtle signals that are often submerged under orders of magnitude by statistical and systematic noise. 
Every technology in this area has to face these problems because, on a cosmic scale, exoplanets are almost completely irrelevant. 
They contribute only infinitesimally to the mass or energy budget of galaxies.
Even in our own solar system major gas-giant planets such as Neptune and Uranus evaded detection until the advent of the modern telescope; the challenge of discovery at light-year distance scales can seem forbidding.

The most successful techniques to reveal exoplanets are {\it indirect} in that they do not witness signals from the planet itself, but rather the planet's influence on its host star. One is the transit method which witnesses a dip in starlight as the planet traverses the observer's line-of-sight to the star. An alternative method is the radial velocity, which records to-and-from perturbations in the velocity of the star, as it is perturbed by the gravitational field of the planet.
The TOLIMAN (Telescope for Orbital Locus Interferometric Monitoring of our Astrometric Neighborhood) program was motivated by the realisation that neither of these methods are suited to answer a fundamental question: are there any potentially habitable exoplanets around the Sun's nearest neighbour twin system -- $\alpha$~Centauri~AB? 
Unfortunately, the transits require an alignment, a very rare event, while radial velocity can find massive  gas-giant planets, but not small rocky exo-Earths  in the habitable zone of the system.

Arguably, a very promising alternative method is the most traditional branch of {\it Astrometry}: the study of deviations in the position of the star in the plane of the celestial sphere that, in this case, are imposed by the motion of the star and the exoplanet around a common center of mass. 
Like all signals in this domain of science, the deviations in position are very small, of the order of one micro-arcsecond. 
To give a sense of scale, for an observer on Earth, this is the angle subtended by a coin held edge-on ($\sim2$\,mm) while standing on the moon. For the specific case we are interested in, the situation is even more interesting. 

$\alpha$~Centauri is a binary star system (thus the A/B), with two stars constantly in motion one around each other. If you could monitor their motion, for example by taking a series of images at different times, you would see the distance between the center of the stars changing as their orbit evolves. After their equivalent of a year this pattern would repeat - thus, by observing the separation between the stars during some time you would detect a periodic signal. This expected signal would be slightly different if you consider the presence or absence of an Earth-like planet as a third element in this system -- and that is the type of perturbation we aim at measuring. One can imagine that at such scales even the smallest deviations in the position of the satellite or thermal effects in its structure and instruments are enough to build up noise in each image, which is orders of magnitude higher than the signal. 
 TOLIMAN has been designed to implement innovative optical principles to deliver a robust estimate of this signal, despite the inevitable presence of many competing random processes and systematic noise. Details can be found in \cite{toliman_proc} and in Sec. \ref{TOLIMAN} of the present work. A critical component for the success of the mission is our ability to extract periodic signals at the milliarcsecond level from a data stream consisting of over a million of images downlinked from the satellite.

The general process to solve this problem has at least two major stages: first, it is necessary to estimate the period of one cycle for the binary star system; then, a more careful analysis of the amplitude deviations at the relevant periods enables the discovery of  additional clues about the presence of the planet. In this chapter we present some first concepts of one of the possible strategies to solve the first stage directly from raw, imaging data.

\begin{figure}
    \centering
    \includegraphics[width=\textwidth]{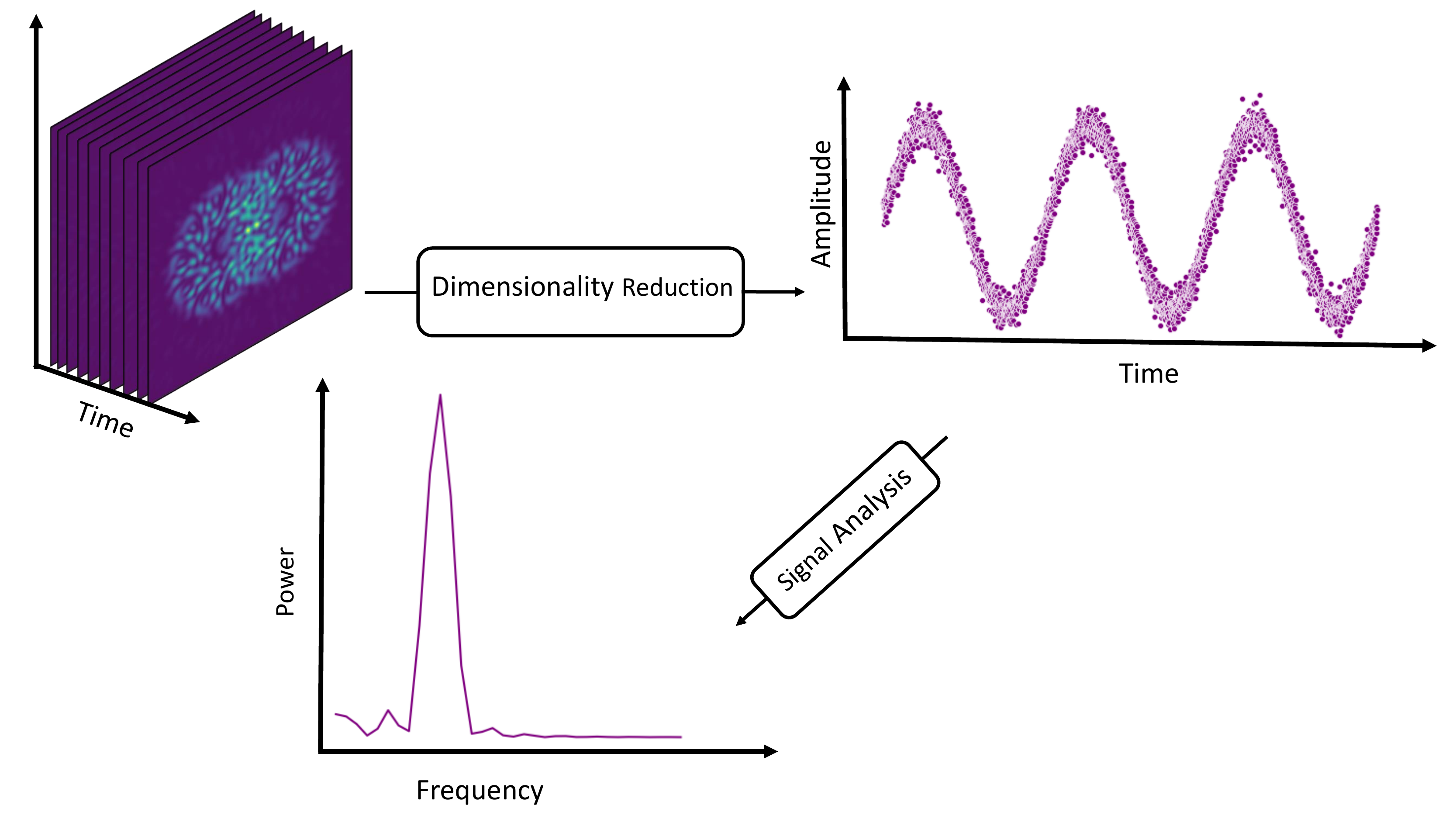}
    \caption{Concept workflow. The underlying concept of the proposed data analysis is based on finding an lower dimensional representation (a compression) of the raw data which preserves periodic signals. Once a suitable representation is found, the effect cause by the presence of the planet can be detected using a time series analysis.}
    \label{fig:workflow}
\end{figure}

Given a series of images of a binary star system as observed by the TOLIMAN mission, our framework uses an unsupervised neural network to learn an abstract (latent), low dimensional representation of the data (the raw images). Here we use a deep convolutional autoencoder \citep{Goodfellow16}. This step reduces the dimensionality of the problem from $256 \times 256$ pixels (size of the images) at each sampling time to $1$ parameter of the latent space, which can then be analyzed as a traditional time series. An overview of the workflow is given in Figure \ref{fig:workflow}. We use simplified simulated versions of the images to be measured by TOLIMAN to show how one can use concepts of neural networks to construct a data analysis pipeline that may be able to extract periodic astrometric signals with an amplitude up to a million times smaller then the pixel size.

 In this chapter, we shall guide the reader through all the modules illustrated in Figure \ref{fig:workflow}. Sec. \ref{Astrometry} gives an overview of the astrometric principles that inspired the  TOLIMAN mission, presented in more details in Sec. \ref{TOLIMAN}. We then show how the simulations were constructed in Sec. \ref{Simulations}, with a brief review of the principles of traditional dimensionality reduction techniques in Sec. \ref{Dimensionality_Reduction}. We introduce basic concepts of deep learning, and how they can be used to learn a meaningful non-linear representation of the input data, in Sec. \ref{Deep Learning}. In Sec. \ref{Architecture} we analyze the architectural choices made to build the deep convolutional autoencoder and  in Sec. \ref{Signal Analysis} the time-series analysis tools, which allow to extract the periodic signal from the data latent space. Once most of the tools are presented, we show the performed experiment and their results in Sec. \ref{Experiment}. We finally draw the conclusions in Chapter in Sec. \ref{Conclusions}.\\
 
\section{Astrometry}
\label{Astrometry}
Before delving in the conceptual diagram of Fig. \ref{fig:workflow}, we want to introduce the astrometric detection field and thus the reasons behind the TOLIMAN satellite architecural choices.
Astrometric detection involves a precise measurement of stellar positions, and is widely regarded as the leading concept presently ready to find earth-mass planets in temperate orbits around nearby sun-like stars \citep[e.g.][]{Shao2009,2017arXiv170701348T}. The principle for detecting a planet using astrometry is the same as that adopted by the hunters of unseen companions of stars \citep[e.g.][]{1844MNRAS...6R.136B} about two hundred years ago. As a planet orbits the star, the latter is tugged in a small circle by reflex motion, thus by careful measurements of the position of the star over time (either in a local or global frame, that must be more stable than the signal produced by the invisible companion). These tiny displacements, imposed on the host star by the gravity of orbiting exoplanets, yield a solution for the planet mass and orbit. Unlike other methods, there are few blind spots, and the signal generated by companions increases with planet-star separation, converse to both radial velocity and transit methods. These unique characteristics make it ideal to probe habitable zones at larger orbital radii. Furthermore, the intrinsic signal (the amplitude of the periodic angular wobble on the sky) is inversely proportional to distance, favoring stars in the immediate neighborhood of the Sun.

However, despite the potential promise, astrometric detection for exoplanetary discovery has not yet entered the mainstream.  The angular excursions induced by habitable-zone Earth-analog planets are small, of order of one micro-arcsecond even for best-case targets, such as Alpha-Centauri. 
Ground based high precision astrometry campaigns must fight the considerable sources of noise, such as the starlight path through the Earth's turbulent atmosphere. Long-baseline optical interferometers have historically delivered precisions better than 100 micro-arcseconds, with a recent resurgence of interest prompted by ESO's GRAVITY instrument \citep{gravity2017} with accuracy an order of magnitude better, which is still not sufficient for Earth-mass planets, however. 
Furthermore, the nearest stars to Earth present a large apparent angular diameter and are correspondingly difficult to observe on long baselines, since they are over-resolved objects, and thus present challenges to the interferometric technique, due to low fringe constrast.
These intrinsic challenges for ground-based astrometric observation have increasing the interest in Space.
Global, large space astrometric surveys over wide angles have proved to be extremely productive delivering fundamental stellar positions, distances and kinematics with the ESA/HIPPARCOS mission \citep{1997ESASP1200}, and its ambitious successor ESA/Gaia \citep{2016A&A...595A...1G, 2018A&A...616A...1G}, which is now measuring billion stars with precisions of the order of $\sim10 \mu$as. Although the Gaia mission expected to deliver a rich harvest of gas giant planets \citep[e.g.][]{2008A&A...482..699C,2018A&A...614A..30R}, in order to detect and study rocky planets in temperate orbits, we need to push detection thresholds down to levels better than $1\mu$as, something that will require dedicated new concepts.

Conventional astrometry approaches measure the position of a star, using a grid of reference nearby objects. This requires relatively large fields of view, since the distance between science targets and sufficiently bright reference objects are of the order of several arcminutes. However, maintaining long-term instrumental stability over such large angles is notoriously challenging. Several interesting missions have been proposed by groups in Europe \citep{2017arXiv170701348T}, the US \citep{sim2008} and China \citep{step2013}, addressing the different concepts to solve this problem with highly stable and continuously monitored spacecrafts and instruments. This poses, however, an additional non-negligible problem: the instrumental cost scales significantly with the field-of-view. Thus it is natural to ask the question if it is possible to obtain micro-arcsecond level measurements for certain targets, like Alpha-Centauri, using much narrower fields-of-view, and thus avoiding the high costs associated to the stability of large field-of-view concepts.

\subsection{Narrow-Angle Astrometry}
\label{NAS}

Our ability to perform narrow field astrometric science ultimately rests on the ability to precisely register the position of the stellar image in each exposure. This meets a fundamental photon noise limit, even with a perfectly stable optical apparatus. Typically any bright nearby star will provide enough photons so that this theoretical limit is not a major problem, requiring only minutes or hours of integration with a telescope of reasonable aperture. However, the critical limitation is not set by photons from the target star. In order to perform the measurement, absolute stability of the image plane sensor is required: something that can only be accomplished with continual monitoring and ongoing calibration. For the practical narrow-field astrometry, registration of the images is performed by simultaneous monitoring of a constellation of background stars, which provide instantaneous information about the exact plate scale and further order deformations. Our astrometric detection error budget is therefore dominated by the accumulation of sufficient counts on these much fainter reference stars that, for a field of view of several arcminutes, are likely to be thousands of times fainter than the target star. The concept underlying the TOLIMAN mission was developed on the principle that it is possible to entirely sidestep this dilemma for the special case of observations of bright binary stars.

Where two bright stars lie close together in the sky, precise monitoring of their separation will deliver the key science with negligible photon noise. In particular, Alpha Centuri is almost ideally tailored for a mission exploiting narrow-angle self-referenced astrometric detection. As our nearest celestial neighbor system, Alpha Cen's pair of solar-analog stars means that habitable-zone exoplanets could be true Earth-twins in year orbits: at the sweet spot for detectability within an attainable mission duration and yielding signals factors of 2--10 times stronger than the next-best systems. The two habitable zones have wide enough orbits to yield good signals, yet not so wide as to require an extended mission lifetime for detection.

\section{TOLIMAN}
\label{TOLIMAN}

The TOLIMAN space telescope \citep{toliman_proc} is a low-cost mission which aims to push the boundaries of astrometric measurements in binary star systems and to enable the detection of Earth-like planets around $\alpha$~Centauri, our closest extra-solar system. The mission is optimised to search for habitable-zone planets that, for $\alpha$~Cen, implies deflections with amplitudes of order of $\sim1 \mu$as over roughly 1-year orbital periods. The detection of such a small astrometric signal has never been reported before in the astronomical literature. 

To accomplish this task with an affordable spacecraft and mission profile, an innovative optical and signal encoding architecture was proposed. It explores and reformulates the idea of a Diffractive Pupil based optical system. 

As originally envisaged, a diffractive pupil telescope would have a set of diffractive features, most simply a regular array of small opaque dots, embedded in the pupil of the instrument \citep[e.g.][]{Guyon2012}. These must be anchored to some element with extreme mechanical stability. The features cause starlight to diffract in the image, essentially forming a pattern whose features are exactly known and stable so long as the diffractive pupil remains stable. For bright sources, this simple concept offers a cunning solution to the key problem that overwhelmingly dominates astrometric error budgets: the stability of the instrument. 

When trying to reference stellar positions at micro-arcsecond scales, a host of small imperfections and mechanical drifts, warps and creep of optical surfaces, generates systematic instabilities that can be orders of magnitude larger than the true signal. Rather than trying to directly contain all these errors, the Diffractive Pupil approach sidesteps them. It creates a new ruler of patterned starlight against which to register positions in the image plane. The cleverness of this approach is that the diffractive grid of starlight suffers identical distortions and aberrations to the signal that is being measured. Therefore, drifts in the optical system cause identical displacements of both the object and the ruler being used to measure it, making data immune to a large class of errors that encompasses other precision relative astrometry approaches.

The opaque dots pupil proposed by \citet{Guyon2012} results in a diffraction pattern where the image plane is populated by a regular grid of sidelobe images diffracted from the bright target star. However, when considering broadband illumination, bandwidth smearing of the starlight will draw each sidelobe into a narrow radial streak or ray. The signal recovery proceeds by registering the location of these rays against the background field stars. Because the diffractive ruler takes the form of long narrow radial rays, positional information recovered must be in the orthogonal ordinate. Therefore, the primary observable consists of the recovery of azimuthal positions of (a rich field of) background stars registered against the nearest diffraction rays. For the TOLIMAN mission, the diffractive pupil formulation described above has two fatal flaws: (1) it relies on background field stars and (2) with its radially smeared ruler it is unable to yield precision measurement of the separation of any binary star. For only a single pair of stars, as is the case of Alpha-Centauri, radial information is essential. Instead, TOLIMAN proposes a novel form of diffractive ruler, which generates fine-featured patterns capable of spanning the required separations between the components of a binary star system. 

\begin{figure} [tb]
   \begin{center}
   \begin{tabular}{cc} %% tabular useful for creating an array of images 
   \includegraphics[height=5.7cm]{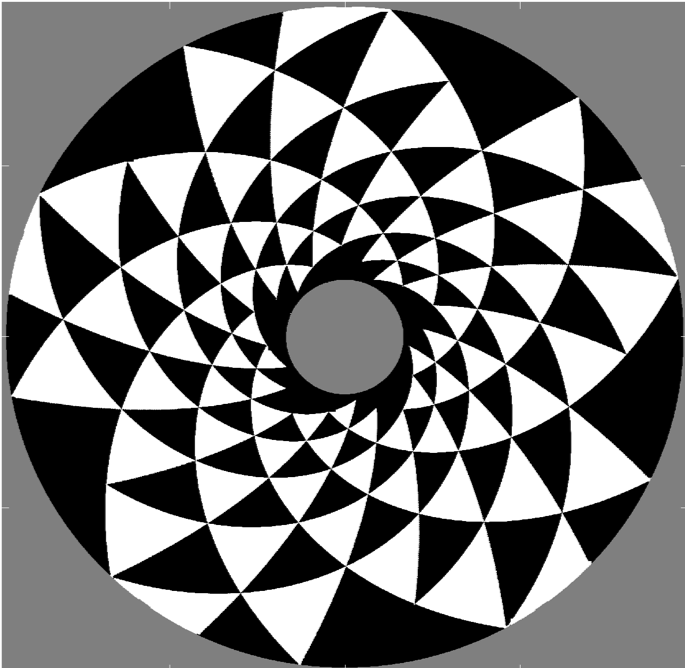}
   \includegraphics[height=5.7cm]{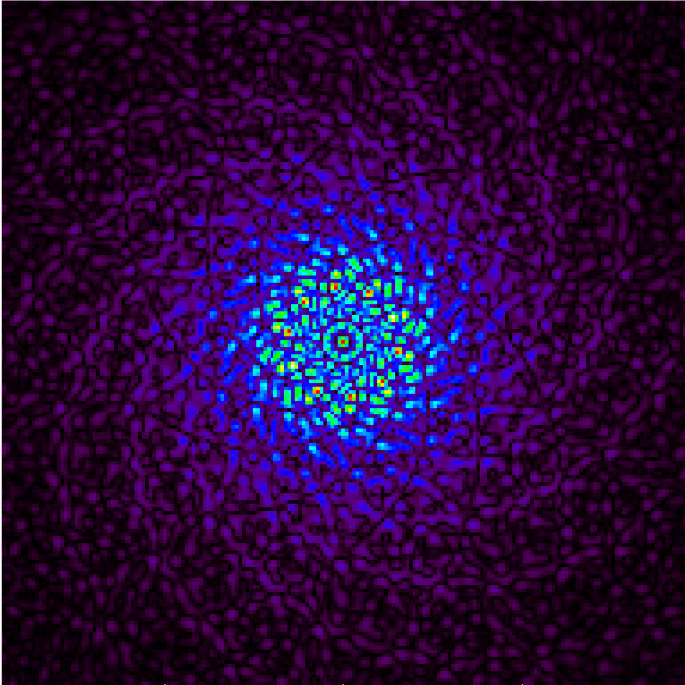}
   \end{tabular}
   \end{center}
   \caption[example] 
%>>>> use \label inside caption to get Fig. number with \ref{}
   { \label{fig:TolPupilPSF} 
Left Panel: a conceptual design pupil for the TOLIMAN mission, with white/black regions indicating discrete phase steps of $0/\pi$. 
Right Panel: the monochromatic PSF generated yields a complex and strongly featured pattern extending from the core, uniformly filling the region with sharp fringes.}
\end{figure}

TOLIMAN requires diffractive pupils capable of creating patterns with a sharp structure extending in the radial direction. Our primary design driver was to find patterns that create a region on the image plane uniformly filled with features that have the highest gradient energy and that occupy the minimum span in dynamic range. Essentially, the former criteria attempt to optimise our ability to accurately register the resulting pattern -- fitting algorithms rely on regions where the image has the strongest slopes or sharp edges. The latter condition is required to spread the starlight preventing saturation of the detector, and spanning the separation of the binary with diffractive features so as to enable the diffractive pupil methodology. Such a design is depicted in Figure~\ref{fig:TolPupilPSF} and is now seen to meet our goal of filling the entire diffractive region, including the core, with sharp structure.
Sharp gradients in the image plane optimise the ability to precisely register such an image.

\subsection{The TOLIMAN Data Challenge}
\label{TOLIMAN_Astrometry}
In its simplest form, extracting the science signal arising from TOLIMAN data requires the exact registration of two overlapping point-spread functions, one for each component of the binary star, in the image sensor plane of the orbiting space telescope. If the separation between these two stellar images can be monitored with sufficient precision, tiny perturbations due to the gravitational tug from an unseen planet can be detected. Given the configuration of the optical system, the scale of the shifts in the image plane are about one millionth of a pixel ($10^{-6}$ pix), thus exquisite stability is required: these motions are only manifest as a sinusoidal perturbation over year timescales. 

Although there are many potential sources of imperfection and error, this first study restricts itself to the most basic and fundamental one, with noise processes arising principally from photon noise and the spatial discretization of the signal. Additional terms, as imperfect spacecraft pointing, jitter and roll stability, will be addressed in future work. For the present study, simulated and laboratory testbed data were created to embody such error terms. 

\begin{figure} [tb]
   \begin{center}
   \begin{tabular}{cc} %% tabular useful for creating an array of images 
   \includegraphics[width = 0.5\textwidth]{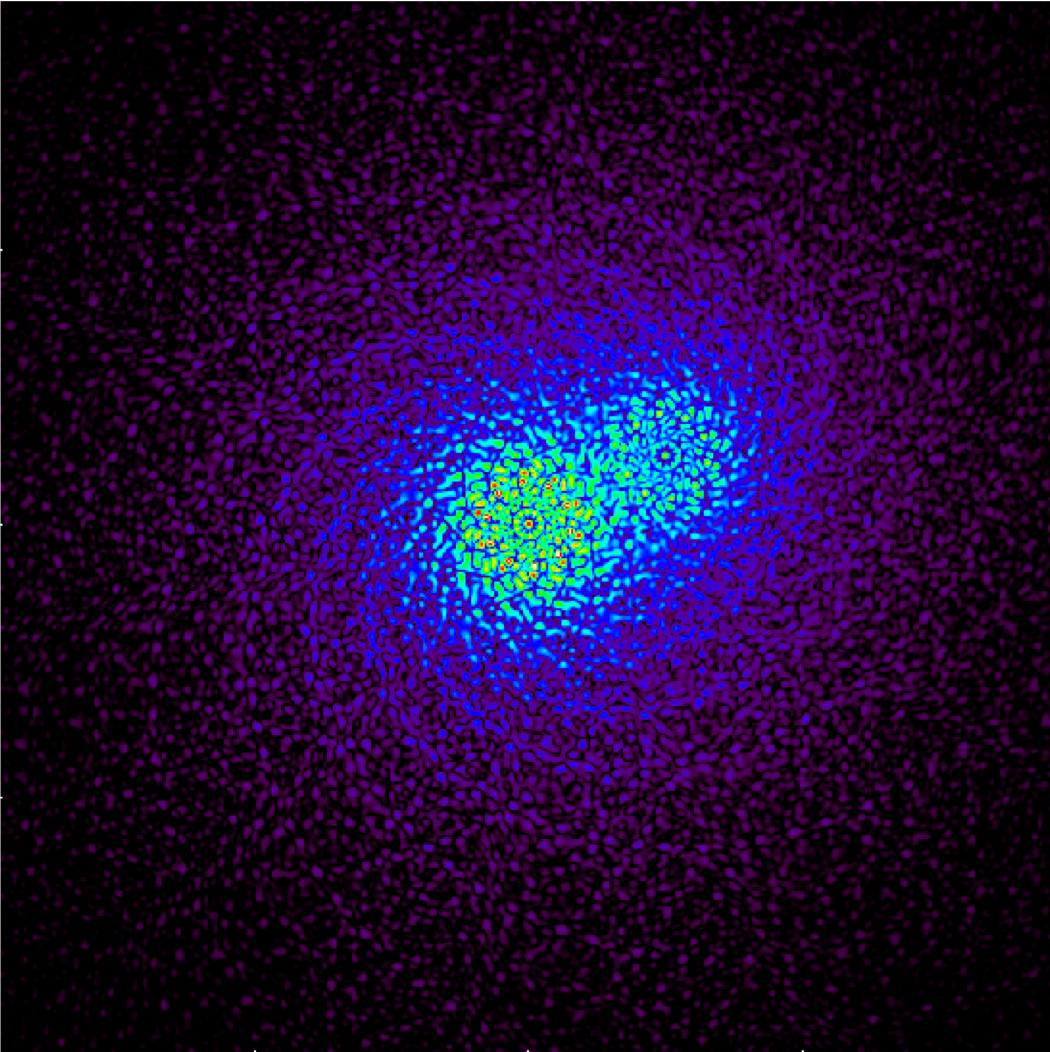}
   \end{tabular}
   \end{center}
   \caption[example] 
%>>>> use \label inside caption to get Fig. number with \ref{}
   { \label{fig:TolBin} 
A simulated binary star as observed with the conceptual design TOLIMAN pupil discussed above. }
\end{figure}

A pictorial illustration of the basic challenge is shown in Figure~\ref{fig:TolBin}: two patterns exist within the frame of data, in this case without the noise terms. High degrees of sharp image structure result in a data for which accurate image registration is possible; however, on the other hand the levels of extreme measurement precision required to obtain the science move this from a relatively routine exercise in image processing (at levels of $10^{-2}$ pixel) to an unsolved problem at signal fidelity levels never yet attempted (at levels of $10^{-6}$ pixel).

\section{Simulations}
\label{Simulations}
We are now ready to describe the simulations of the TOLIMAN data, i.e. the inputs of our conceptual workflow shown in Fig. \ref{fig:workflow}. These simulations were necessary given that no testbed has been build yet to accurately reproduce the TOLIMAN data.
This section, thus, describes the formalism and necessary steps to produce a mock data set that mimics the precision level required by the TOLIMAN mission. The first challenge in developing a method capable of extracting a signal as small as one millionth of a pixel is to develop a computational model capable of emulating such signal under varying conditions of noise. Although injecting a signal into an image may seem a rather trivial task, common approaches fall short when pushed to the limits of precision required by the TOLIMAN mission, often resulting in large computational cost. The traditional simulation approach consists of generating a super-sampled Point Spread Function (PSF). Since stars can be considered point sources, to simulate a stellar field as would appear on the detector, we simply need to shift and downsample that PSF in the sensor grid. Thus, by assigning it to either random or specified positions within the image and repeating the procedure for many different point sources, we can recreate a stellar field. While this can be made computationally efficient today using the widespread GPU accelerators, such traditional methods unfortunately introduce errors orders of magnitude greater than the signal we expect to measure, thus requiring alternative approaches to the generation of the mock data.

\subsection{The Fast Fourier Transform}
The Fast Fourier Transform (FFT) has long been used as an optical simulator since it performs the same operation numerically as a focusing mirror or lens does optically. An input image will undergo a transformation from a spatial representation to a frequency representation when observed at the focal plane. The Fast Fourier Transform operates on a digitised representation of the input with an $O(n.log(n))$ computational complexity, making it a corner stone in basic computations of optical systems. In this section, presenting the basic underpinnings, we detail how one can use FFT's to create images of stellar fields with the injection of arbitrarily sized positional information.

Generating these PSF's is conceptually straightforward, requiring only the representation of the electric field at the aperture $ E(x, y, \lambda) = A(x, y) e^{i \theta(x, y, \lambda)} $ as its amplitude $A(x, y)$ and phase $\theta(x, y)$and combine these terms into a complex array. The PSF in the $(u, v)$ focal plane is then found by taking the power of the resultant FFT of the complex array.

\begin{equation} \label{FFT}
    PSF(u, v) = |\mathcal{F}\{E(x, y, \lambda)\}|^2 = |\mathcal{F}\{A(x, y) e^{i \theta(x, y, \lambda)}\}|^2
\end{equation}

Positional information can then be injected through applying a linear gradient to the phase $\theta$. An Optical Path Difference (OPD) is introduced across the aperture by any source off-axis from the normal of the telescope pointing. Easily calculated through the angular offset from the normal, the OPD simply translates into phase as a function of the observation wavelength. 

\begin{equation}
    \theta_{slope}(x, y, \lambda) = \frac{2\pi}{\lambda}OPD(x, y)
\end{equation}

Having this mathematical representation of on-sky position to telescope response allows for arbitrary signal sizes to be introduced to any stellar objects. Other natural or designed phase perturbations like optical aberrations (coma, astigmatism, etc) or devices like the TOLIMAN diffractive pupil can easily be represented and added to the other phase sources. Optical aberrations are not explored in this work but the principles underpinning their simulation follows simply from this work. Other phase devices like the TOLIMAN diffractive pupil $\theta_{Pupil}(x, y, \lambda)$ follow the same general idea. Formulated as a mirror with 'steps' cut in, we take the height of each step $h(x, y)$ and translate to phase by taking the OPD as twice the height of the step.

\begin{equation}
    \theta_{Pupil} (x, y, \lambda) = \frac{2\pi}{\lambda}2h(x, y)
\end{equation}

The total phase $\theta$ is then a linear combination of these effects. Taking the field amplitude $A(x, y)$ as unity for all non-masked regions of the aperture gives the full description of the electric field $E(x, y, \lambda)$. 

Having formulated the electric field response to the system, we must introduce a complete description of the optical architecture. This is described by a handful of parameters: aperture diameter $D$, effective focal length $fl$ and pixel size $d_{pix}$. Desiring computational efficiency through the inclusion of our optical system, we define some value $N_{out}$ to be the size of the array which we pass to the FFT. This is the primary driver behind the computation cost. Using this value and the described parameters, the size of the array $N_{E}$ representing our electric field $E(x, y, \lambda)$ can be found. Note all arrays are taken to be of size $N\times N$. 
These two values necessarily differ as a way to encode optical parameters without focal plane interpolation. The ratio between $N_{out}$ and $N_E$ determines sampling in the focal plane matching that of our system.

\begin{equation}
    N_E = N_{out} \frac{d_{pix} \times \lambda}{D \times fl}
\end{equation}

Embedding this array representing the electric field into an $N_{out}$ sized array, we use equation \ref{FFT} to generate a PSF that requires no interpolation and can have positional signals of any size injected, limited only by floating point precision of course. Further details and descriptions of these processes can be found in \cite{doi:10.1080/00107514.2016.1248491}.

\subsection{Generating Data}

With the tools to simulate PSF's through our optical system we must now generate a data set. By adding basic noise processes, stellar spectrum and astrometric signals we can create a comprehensive set of images that can be used to test the recovery and reconstruction abilities of all the data driven techniques described in this chapter. Here a balance must be struck, generating a truly comprehensive data set for the TOLIMAN mission is simply intractable. With a full signal period of order one year, any data set must present the basic challenges of the mission in an efficient way. Here we examine choices such as number of wavelengths, stars and images to simulate, along with the included noise processes.

One of the first things to consider is the size of the data set, the total number of images produced. The TOLIMAN signal is introduced to the $\alpha$ Cen system through the gravitational tug of an orbiting planet and so our signal is sinusoidal by nature. The orbital period that we are searching for is of order of a single year, and so producing a 'frame by frame' data set would be computationally intractable. Consequently we need to generate each 'image' as a representation of a collection of multiple from the actual telescope. We chose to represent three full signal cycles over $1000$ images, with each image representing approximately a full day. 

\begin{figure}
    \centering
    \includegraphics[width=0.9\textwidth]{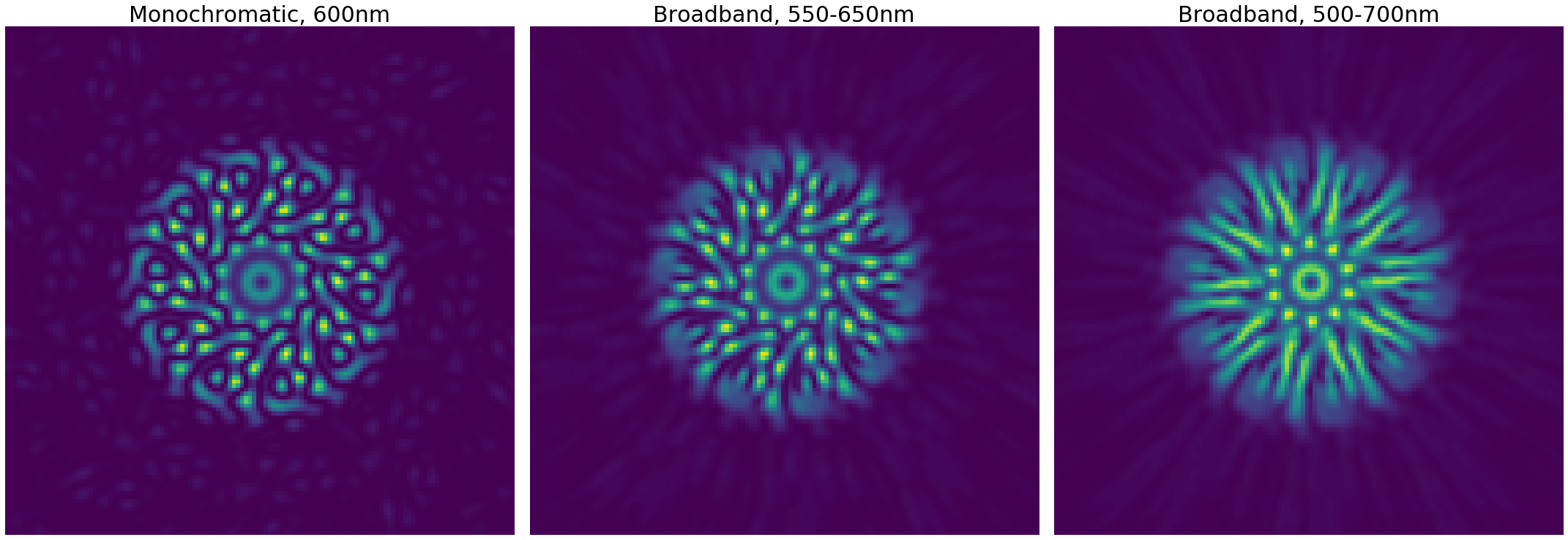}
    \caption{TOLIMAN PSF at different bandwidths. Left: Monochromatic 600nm. Centre: 550-600nm (best resembles actual mission). Right: 500-700nm}
    \label{fig:broadband}
\end{figure}

Observing in the visible spectrum over a 100nm bandwidth, the choice spectral resolution is important. The wavelength dependence of the PSF demands that the image at each wavelength be computed individually. To represent the real world as closely as possible, a spectral resolution of 1nm was chosen for a number of reasons:
1) firstly the Toliman PSF is spread over many diffraction limits (10$\lambda$/D) so at the outer reaches bandwidth smearing begins to have a large effect on the PSF shape (for an example see figure \ref{fig:broadband}); 2) secondly, by choosing to maintain the stellar alignment on the detector constant and keeping one of the stars stationary, we are able to massively reduce the number of PSFs we must compute. A stationary star only requires calculation of a single broadband PSF. For the moving star, since the TOLIMAN signal is sinusoidal by nature and the stellar alignment is kept stationary, we only need to calculate the PSFs for a single signal cycle. The result is a large overhead for small simulations but with the benefit of being able to produce large and accurate simulations efficiently.

Given our spectral resolution, we can use one of the many libraries available to generate spectra that reflect the true stellar parameters for each star. These libraries access existing stellar databases and recreate synthetic spectra for a host of variable stellar parameters such a effective temperature, metalicity and observational flux. We used \emph{Pysynphot} [\cite{2013ascl.soft03023S}] to generate stellar spectra and fluxes for our system. This package uses models built from HST observations across the HR diagram to simulate atmospheric emissions from different stars. Taking the relative fluxes and total photon counts output from this system, we can scale each monochromatic PSF by its relative power to recreate accurate PSFs.

While real data will feature many varied noise processes, here we only consider two noise sources: photon and detector noise. These are dictated by Poisson and Gaussian statistics respectively. Detector noise is primarily driven by the random thermal fluctuations of the discrete electrons that carry the signal through the detector. With available modern low-noise sensors, this noise is not expected to limit the extraction of the signal since it averages out to some constant value over many frames. The addition of even modest levels of this noise also serves a separate motivation: to allow for a smoother error space. This helps numerical algorithms converge faster as fine structure in the gradients are rounded and the algorithms are able to follow a smooth descent to the optimum. On the other hand photon noise is an important processes that must be examined. Arising from the discrete nature of photons, this noise is simulated at each pixel by drawing from the Poisson distribution whose mean is dictated by the PSF. When performing image registration of small signals such as those anticipated in the TOLIMAN mission, the total number of photons that arrive in each image becomes an important factor. As shown in \cite{Guyon2012}, there is a fundamental relationship between the number of photons received and the positional information carried by those photons. With insufficient photons, signals can not be extracted. 

Simulations proceeded with the production of a comprehensive batch of noisy image data sets, with sinusoidal signals in separation of the binary star injected with decreasing amplitude to mimic increasingly more challenging planets, up to the limiting deflection of one-millionth of a pixel. These simulations closely resemble the expected response of the TOLIMAN optical system to the observation of the $\alpha$~Cen system and were used to build and train machine learning algorithms.

\section{Dimensionality Reduction}
\label{Dimensionality_Reduction}
As it can be seen in the conceptual diagram in Fig. \ref{fig:workflow}, the first and the most important step in the proposed data analysis scheme is to apply some transformation to the raw data produced by the instrument to allow us to unveil the periodic changes in the images through time. This transformation can be seen as a dimensionality reduction, a compression of the imaging data into a smaller dimensional space that preserves periodic signals that may exist in the data. 

Dimensionality reduction  \citep[e.g.][]{pearson1901}, that is, of representing data in a different space than that in which it can naturally be observed, is a set of techniques that try to transform data from one representation to a lower dimensional one with the lowest information loss. Reducing dimensionality of data with minimal information loss is important for feature extraction, compact coding and computational efficiency, to eliminate redundancies and enforce constraints. In particular image compression techniques try to take advantage of statistical properties of the images in order to reduce their computational footprint.
One of the simplest and most widely used approaches of dimensionality reduction is the Principal Components Analysis (PCA) \citep{dunteman1989}. This approach consists in applying a linear projection of the original data on a set of orthogonal axes (the principal components), built to recover the maximum amount of information contained in the original data with as few coefficients as possible. In practice, PCA can be computed by performing a singular value decomposition of the data, contained in a matrix $X$.
Each datapoint can then be reconstructed by a linear combination of the basis elements: $X\approx DA$, where $D$ is a matrix containing the principal components, and $A$ a matrix containing the coefficients used in their linear combination when reconstructing the data. Another broad class of dimensionality reduction methods, closer in heuristic to using a neural network to build the new representation space, is that of dictionary learning (DL). Instead of using PCA to select the new basis of representation $D$, one can instead \emph{learn} it from the data itself. Much like in deep learning approaches, dictionary learning relies on the choice of a loss function $l$ to quantify the difference between input data and its reconstruction. Learning the representation then amounts to solving the following optimization problem:

\begin{equation}
    \min_{D,A} l(X, DA).
\end{equation}
Depending on the desired properties of the representation to be learned, one can further add \emph{constraints} to either the dictionary $D$ or the coefficients $A$. A great many flavours of dictionary learning exist, depending on the constraints selected. One of the most widely used is the addition of a \emph{sparsity} constraint on $A$ \cite{mairal2009}. In practice, the sparsity constraint is often obtained by adding an $l_1$ term to the cost function:

\begin{equation}
    \min_{D,A} l(X, DA) + \lambda \|A\|_1.
\end{equation}
Several other constraints exist, and often lead to the resulting dictionary learning approach having its own name: non-negative matrix factorization \citep{lee2001} when using positivity constraints, sparse PCA \citep{d2005direct} when the sparsity constraint is instead imposed on the dictionary, independent components analysis \citep{hyvarinen2000} when imposing statistical independence between the components
Both PCA and DL were studied in the development of the work described in this chapeter to compress the images, and a period consistent with the one of the signal injected in images could be found in the produced lower dimensional representations up to a signal amplitude of $10^-4$. These techniques were not capable to detect astrometric signals with amplitudes at the order of $10^-6$ times smaller than the pixel size. However, their application showed us that through data driven techniques, the images could be transformed into a lower dimensionality space while preserving the temporal signal structure and thus that the challenging $\mu$as-level signals could perhaps still be recovered with the use of other techniques, such as Deep Learning. The next section makes a small, but self-contained, introduction to these other techniques.

\section{Deep Learning}
\label{Deep Learning}
In this section we will review all the concepts underpinning Deep Learning needed to understand the inner workings of the deep convolutional autoencoder used in this work to create a lower dimensional representation of the TOLIMAN simulated data. The main advantage of these techniques over classical dimensionality reduction, is that the layered structures of Deep Neural Networks (DNNs) can encode an input representation with increasing levels of abstraction in successive layers \citep{lecun2015, Goodfellow16}. For such reasons, in the last decade, Deep Learning has been successfully applied across a wide range of applications including computer vision, speech recognition, bioinformatics and astroinformatics.

In this work we make use of two classes of Neural Networks: fully connected and convolutional. Fully connected Neural Networks, also simply known as Neural Networks, can be used to approximate any nonlinear functional relation between a set of inputs and outputs \citep{cybenko1989}. Each layer of a neural network transforms a vector of inputs $x \in R^{N}$ as follows:
\begin{equation}
    \hat{y} = f(Wx + b), 
\end{equation}
where $W \in R^{(N \times K)}$ is a matrix of weights, $b \in R^K$ is a bias term, and the nonlinear activation function $f: R \rightarrow R$ is applied componentwise. The bias term shifts the baseline activation function input away from zero, providing richer behavior for modeling the functional relationship between the input and output variables. In networks with multiple layers, the output of each layer is connected to the input of the following one
\begin{equation}
     \hat{y_{l + 1}} = f_{l + 1}(W_l h_l + b_l) = f_{l + 1}(W_l f_l(...(f_1(W_0x + b_0) + b_l)
\end{equation}
where $h_l$ is the hidden layer or feature vector of layer $l$. The input is processed through all the layers until the output of the network $\hat{y}$ is reached.

The parameters (weights and biases) of the network are selected to minimize a \emph{loss function}, such as a mean square error, summarizing the difference between the network output and a desired or observed target value $y$.  Stochastic gradient descent (SGD) is a common optimization process for neural networks:  at each stage of training, the network parameters are updated by a small vector proportional to the gradient of the loss function with respect to those parameters.  This is straightforward for the output layer; weights and biases in overlying layers can be efficiently calculated through successive applications of the chain rule for derivatives, in a process called \emph{backpropagation}.  The use of derivative information for efficient network training requires that the loss function be smooth.

Convolutional Neural Networks (CNNs) differ from fully connected neural nets only in that their architecture exploits the localized structure of images to reduce the number of network parameters needed.  Instead of connecting each neuron in a layer to every other neuron in the next layer, the connection structure of CNN layers is sparse, and parameters are shared across a layer to enforce translation invariance of features extracted on each scale across the image.  Three types of layers are typically used: 1) Convolutional Layers, 2) Pooling Layers and 3) Fully-Connected Layers. In the following, we will analyze in detail the architecture and inner workings of each one of them.

\subsection{Convolutional Layer}
\label{Convolutional_Layer}
The Convolutional Layer is the most computationally intensive part of a CNN architecture; its parameters consist of a set of learnable filters. Every filter, also know as a \emph{kernel}, is spatially small (usual sizes are $3 \times 3$, $5 \times 5$ and so on, where 3 and 5 are sizes in number of pixels), but includes weights for each channel of its input.  For the first layer, these channels are the data channels (for example R, G, B in a three-channel image).  In subsequent layers, each channel corresponds to the output of a single kernel from the previous layer. During the forward pass, each kernel slides across the spatial dimensions of the input, computing the dot product between itself and the part of the input volume that it encompasses (\emph{convolution}). As the kernel slides, it produces a bi-dimensional activation map that encodes the responses of the kernel at every spatial position. The content of the activation map at each location is a direct response to some visual feature present in the image to which the kernel is sensitive, such as an edge or a color. Each convolutional layer employs multiple different filters, producing a set of activation maps that are stacked along the depth to produce a multi-channel output. Due to the limited size of the filters, neurons are not connected to the full extent of the input volume but only to a small region (the \emph{receptive field}). The connections are thus local in space (width and height of the input), but are always fully connected in depth (i.e. across learned/extracted features).

The structure of the output volume of a convolutional layer is controlled by three hyper-parameters:
\begin{itemize}
    \item \emph{Depth}: the number of filters learned in the layer;
    \item \emph{Stride}: the number of pixel the filter is shifted along the spatial dimensions of the input volume. It is usually set to one, but can be set to higher values, depending on the image geometry, to achieve less redundancy in the output volume. The stride controls the spatial dimensions of the output volume;
    \item \emph{Padding} or \emph{zero-padding}: the width in pixels of a spatial region on the borders of the output that is filled with zeros. It controls the spatial dimensions of the output volume and can be used to preserve the spatial dimensions through the layer.
\end{itemize}
Finally, to ensure that each kernel is learning a single feature that has a consistent interpretation across the spatial extent of the input, all neurons in the same depth slice share the same weights and biases, irrespective of where across the extent of the input they are applied.  Thus the action of each filter in the forward pass becomes a discrete convolution of a single set of kernel weights with the input.

A convolutional layer acts to encode its inputs into a latent space spanned by the features it learns.  However, the autoencoder architecture we will consider in later sections also involves a transformation from a learned latent space back into the image domain.  Thus, while convolutional layers typically decrease the spatial extent of their inputs, we will also need \emph{deconvolutional layers} which increase them, recombining a potentially large number of learned features into a flat image.
Mathematically both convolutional and deconvolutional layers can be summarized as
\begin{equation}
    l^{h} = f\Bigg(  \sum_{i \in L} x^{i} \otimes w^{h} + b^{h} \Bigg)
\end{equation}
where $l^{h}$ is the latent representation of the $h$-th activation map of the current layer, $f$ is the activation function, and $x^{i}$ is the $i$-th activation map of the $L$-feature activation of the previous layer in the network (or the $l$-th channel of an $L$-channel image in the case of the first convolutional layer after the input image). $w^{h}$ and $b^{h}$ are, respectively, the weights and biases of the $h$-th activation map (shared by all neurons of the map) of the current layer. Given that $x^{i}$ has size $m \times m$ and the filters have size $k \times k$, a convolutional layer produces an output feature map with shape $(m - k + 1) \times (m - k + 1)$, thus reducing the size of the input. A de-convolutional layer outputs a feature map with shape $(m + k - 1) \times (m + k - 1)$, thus increasing the size of the input.

\subsection{Pooling Layer}
\label{Pooling_Layer}
The architectural function of a Pooling Layer is to reduce the spatial size of the representation, which reduces the number of parameters, lightens the computational load, and mitigates overfitting. The pooling operation is carried independently on each input feature, leaving the number of input features unchanged. Different criteria in the literature exist to perform the pooling operation, including \emph{max}, \emph{average} and \emph{L2-norm pooling}; max pooling is the most commonly used. There are also \emph{un-pooling} layers to desegregate and expand activation maps in transformations back towards the image domain.

\begin{figure}
\centering
\includegraphics[width=0.8\textwidth]{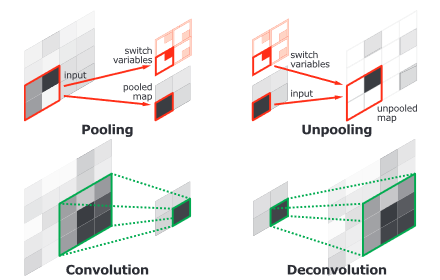}
\caption{Illustration of max-pooling, unpooling, convolution and deconvolution layers.~\cite{NohHH15}}
\label{fig:Pooling}
\end{figure}

A max-pooling layer pools features by computing the maximum within the feature map and outputs a feature map with reduced size, according to the chosen size of the pooling kernel. To perform a successive un-pooling, the max-pooling layer also records a set of switch variables which describe the positional information relative to the pooled features. The un-pooling layer restores the max-pooled features into the correct position specified by the relative switch variable values. The combination of max-pooling and un-pooling layers is thus able to retain both the image magnitude (answering the "what" question) and the positional information (the "where" question). Figure \ref{fig:Pooling}~\citep{NohHH15} shows some stylised representations of convolution - de-convolution and pooling - un-pooling operations.

\subsection{Deep Convolutional Autoencoder}
\label{Auto Encoder}
An Auto-Encoder model (AE) is a neural network composed by an encoder and decoder part; the encoder $f: X \rightarrow H$ transforms the input image into a lower dimensional representation (the \emph{latent space}), while the decoder $g: H \rightarrow X$ tries to reconstruct the original input image from this representation. By constraining the latent space to be of lower dimension than the original input data, we can force the autoencoder to capture the most important features of the input data in order to reproduce it successfully. This type of restriction can be used for feature extraction and for dimensionality reduction.

During the learning process, network parameters are adjusted to minimize a loss function
\begin{equation}
    L(x, g(f(x)))
\end{equation}
that encodes the difference between the input $x$ and its reconstruction $g(f(x))$. As for the the NNs discussed in Sec. \ref{Deep Learning}, $L$ must be smooth in order to use gradient-based minimization algorithms such as SGD. If $L$ is chosen to be linear, the auto-encoder performs a dimensionality reduction similar to Principal Component Analysis (PCA); in fact, the latent space $h$ ends up to be the principal subspace of the input data. If, instead, $L$ is non linear the auto-encoder can learn much complex representation.\\
Generally autoencoders are built by two shallow fully connected NNs joined through a lower-dimensional latent space. A CAE (Convolutional AutoEncoder), instead, contains, in the encoder part, a stack of convolutional and max pooling layers before the fully connected layer and, in the decoder part, a stack of up-sampling and de-convolutional layers after the fully connected layer. It has been shown~\citep{ZhaoMGL15} that CAE are better suited, with respect to AE, for image processing and reconstruction tasks, due to the full utilisation of the CNNs capacity to extract a hierarchical set of features from the images. These have been proven to show a better performance over shallow neural networks when working with noisy or complex images. Moreover the combination of a convolutional and max-pooling layer allows the higher-layers representations to be invariant to small rotations and translations thus helping with the TOLIMAN satellite inevitable jitters and translations.\\
In recent years AEs have been applied to solve a wide range of problems in the Astrophysical context; to model the Point Spread Function of Wide Filed Small Aperture Telescope \citep{10.1093/mnras/staa319}, to uncover and separate the faint cosmological signal from the epoch of reionization \citep{10.1093/mnras/stz582}, to classify galaxies Spectral Energy Distributions \citep{1705.05620}, to identify Strong Lenses candidates in the simulated data of the Euclid Space Telescope \citep{1911.04320} and to solve the Star - Galaxy classification problem \citep{HAORAN2017282}. Moreover in the fields of Computer Vision and Image Processing, AEs have been successfully used to recover structured signals from natural images \citep{1508.04065}, for image compression \cite{1802.09371}, \citep{1703.00395}, \citep{1607.05006}, \citep{1608.05148}, \citep{1601.06759}, achieving compressing performances similar or better than the JPEG 2000.\\ 
Encouraged by the results obtained in literature in lossless image compression and signal recovery from images through AEs, we decided to develop our custom CAE architecture to recover the astrometric signal from the TOLIMAN simulation images. Sec. \ref{Architecture} contains an in-detail description of the architectural design, given the peculiar nature of the scientific problem.

\section{Model Architecture}
\label{Architecture}
In this section we take implement knowledge detailed in Sec. \ref{Deep Learning} to build the actual CAE architecture that compresses the TOLIMAN simulated images into a latent space that showed a periodic trend with time. Some of the architectural choices came from our knowledge of the physical problem, some from the expected behaviour of the network, and others were discovered on a trial-and-error basis.

\begin{figure}
    \centering
    \includegraphics[width=\textwidth]{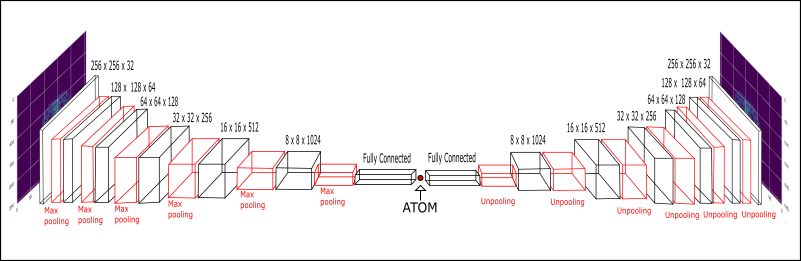}
    \caption{Architecture of the Deep Convolutional Auto-Encoder}
    \label{fig:CAE_architecture}
\end{figure}

Figure \ref{fig:CAE_architecture} shows the overall architecture of the CAE. Each convolutional and de-convolutional layer is followed by an \emph{Exponential Linear Unit} (ELU) activation function. It has been shown by \cite{ClevertUH15} that this function is able to capture the degree of presence of particular phenomena (the signal) and not its absence, thus creating in the network a complex weight space, chains of connections specialised in solving particular tasks (like encoding the signal). Moreover, since ELU may have negative values, it pushes the mean of the activations closer to zero. Having mean activations closer to zero causes faster learning and convergence.
Said that, ELU is very similar to RELU, except for negative input values. In fact ELU becomes smooth slowly until its output equals $-\alpha$ where RELU sharply smooths.

\begin{figure} [tb]
   \begin{center}
   \begin{tabular}{cc} %% tabular useful for creating an array of images 
   \includegraphics[width=0.5\textwidth]{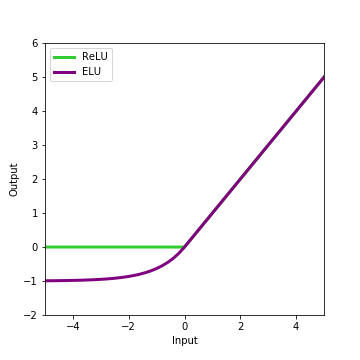}
   \end{tabular}
   \end{center}
   \caption{Comparison Between Rectified Linear Unit and Exponential Linear Unit activation functions.}
   
\end{figure}
For negative activations, RELU's gradient will be 0 and this may prevent the network weights to be adjusted during descent. This means that all the affected neurons going into that state will stop, responding to variations in input (being the gradient 0 there is no input that can make them change, they have reached a local minima from which they are unable to escape). This is called dying RELU problem. Apart from the described computational problem, RELU is less responsive to negative activations, something that may harm the signal reconstruction in the latent space for all images where the binaries separation is smaller than their mean separation in the training set. For all these reasons, we chose ELU as the activaction function of all hidden layers.\\
The Network  latent space was chosen to be uni-dimensional (represented by the single ATOM in Figure \ref{fig:CAE_architecture}), for the following reasons: (i) when a higher dimensional latent space was used, the Pearson correlation coefficient between the latent variables was found to be compatible with a value of  $1.0$; (ii) given that the separation of the star's PSFs is radial, and, given that the only varying feature in the images is the signal, it seems reasonable to think that the only information the network needs to recover from the latent space in order to decode, and thus reproduce, the images is the signal itself. The remaining constant information (pixel luminosity and image geometry) can be stored in the network weights.
In literature, Deep Neural Networks tend to employ two types of loss functions: the Mean Square Error (MSE) and entropy-based loss function like cross-entropy or binary-entropy or the kullback leibler divergence. Although all types of loss functions have explicit probabilistic interpretations, MSE is estimating the mean of any distribution, while the entropy based functions try to maximize the likelihood of a multinomial distribution, they differentiate in their application field. The latter type, with a logistic output, tends to heavily penalize wrong class predictions and thus are specifically suited to work in classification tasks where the decision boundary is large. The first (MSE) is very forgiving on mis-classifications, but is well suited to handle regression problems, where the distance between two predicted values is small. Since our scope is dimensionality reduction, i.e. a regression problem, the MSE was chosen. To test the quality of the image reconstruction, we computed the mean MSE (MMSE) and the mean Structural Similarity Index (MSSI), \cite{Wang04imagequality}, between all the available images and their reconstructions. the SSI models any image distortion as a combination of three factors: correlation loss, luminance and contrast distortions. When comparing two images, the estimator takes into account the mean luminance difference between the two images, the closeness of their contrast and their correlation coefficient.
The number of layers, filters and other layer parameters were  heuristically chosen through a trial-and-error campaign.

\section{Signal Analysis}
\label{Signal Analysis}
As the reader can see from the workflow figure (Fig. \ref{fig:workflow}), the second step in the proposed pipeline is to perform the Signal Analysis in order to unveil periodic trends in time. For these reasons, in this section we present an overview of the chosen method, for instance the Lomb Scargle Periodogram technique, explaining the pre-processing steps performed to compare the atom time series (see Sec. \ref{Architecture} and \citet{VanderPlas2018} for details.)

\subsection{The Lomb-Scargle Periodogram}
\label{Lomb-Scargle}

The most commonly used tool for period searching in irregular cadence astronomical light curves is the Lomb-Scargle periodogram \citep[LSP,][with 4000 citations]{Lomb1976,Scargle82} that assumes a sinusoidal periodic behavior. It is a generalization of the Schuster periodogram in Fourier analysis 
\begin{equation}
P_s(f) = \frac{1}{N}\left\lVert\sum_{n = 1}^{N}g_{n}e^{-2\pi ift_n}\right\rVert^2,
\end{equation}
but for irregularly cadences. 
The LSP stands out as a robust procedure to build a power spectrum in order to detect periodic components in unevenly sampled datasets. 
In the uniform sampling regime, the Schuster periodogram encodes  all of the relevant frequency information present in the data. This definition can be  generalized to the non-uniform case, which is the scenario we explore here. 
It follows that the generalized form of the periodogram addressed by \citet{Scargle82} takes the form:

\begin{equation}
  P_{LS}(f) =
  \frac{1}{2} \Bigg\{ 
 \frac{\left\lVert\sum_{n = 1}^{N} g_n \cos(2\pi f [t_n-\tau])\right\rVert^2} {
  \sum_{n = 1}^{N} \cos^2(2\pi f [t_n-\tau])} 
   +  \frac{\left\rVert\sum_{n = 1}^{N} g_n \sin(2\pi f [t_n-\tau])\right\rVert^2}
  {\sum_{n = 1}^{N} \sin^2(2\pi f [t_n-\tau])}  \Bigg\},
\end{equation}
where $\tau$ is specified for each $f$ to ensure time-shift invariance:
\begin{equation}
  \tau = \frac{1}{4\pi f}\tan^{-1}\Bigg(
  \frac{\sum_n \sin(4\pi f t_n)}{\sum_n \cos(4\pi f t_n)}\Bigg).
\end{equation}
This modified periodogram differs from the classical periodogram only to
the extent that the denominators $\sum_n \sin^2(2\pi f t_n)$ and
$\sum_n \cos^2(2\pi f t_n)$ differ from $N/2$, which is the expected value of
each of these quantities in the limit of complete phase sampling at each
frequency.

\subsection{Atom Time Series Analysis}
To estimate the period of the atom time series, we used the Lomb Scargle Periodogram and to validate the goodness of the period estimation, we employed the following metrics:
\begin{itemize}
    \item False Alarm Probability (FAP): encodes the probability of measuring a peak of a given height (or higher) conditioned on the assumption that the data consists of Gaussian noise with no periodic component;
    \item Full Width at Half Maximum (FWHM): this is an expression of the extent of a function given by the difference between the two extreme values of the independent variable at which the dependent variable is equal to half of its maximum value. Treating the FWHM as an error measure, we derived an error on the period through the following expression:
    \begin{equation}
     P = \frac{1}{f(peak)}
    \end{equation}
    \begin{equation}
     \Delta P = \frac{1}{f(peak)^2} \times \Bigg[ f\Big(peak + \frac{FWHM}{2}\Big) - f\Big(peak
     \frac{FWHM}{2}\Big) \Bigg]
    \end{equation}
\end{itemize}
\noindent where $peak$ stands for the peak of the power spectrum and $f(peak)$ its relative frequency.

In order to compare the atom time series and the signal, we standardized both of them, i.e. with subtracted to both time series their mean values and divided by their standard deviations. This preprocessing step was needed due to the Network inability to perfectly recover the signal amplitude in the latent space.

\section{Experiments and Results}
\label{Experiment}
This section describes all the experiments performed with the CAE to compress the images to a lower dimensional representation, showing a periodic trend in time, i.e. a latent space that preserved the signal, analysing the compressed representation in search of a periodic signal in time.

Before deploying the model on the simulations containing the signal with an amplitude a factor of $10^{-6}$ smaller than the pixel dimension (for details on the simulations see Sec. \ref{Simulations}) and the realistic binaries PSFs flux ratio, the Network encoding capabilities were tested on images containing signals with amplitudes respectively $10^{-2}$, $10^{-3}$, $10^{-4}$, $10^{-5}$ smaller than the pixel dimensions, equal flux PSFs (the binaries PSFs presented the same flux) and an image peak value of $10^9$ photons and photon noise arising from the Poisson statistics.
Due to the absence of any realistic noise components (jitter, rotations, aberrations etc.),
 each image was cropped with a $256 \times 256$ pixels window centered around the image barycenter. This preprocessing was needed in order to eliminate any spurious shift in the image pixels that could have compromised any training attempt capable of extracting the signal from the images. In fact both the max-pooling and convolutional layers (see Sec. \ref{Deep Learning}) are not shift invariant and, as clearly shown in \cite{zhang}, the presence of a shift can completely change the outcome of these operations in an unpredictable way. Each dataset thus contains $1095$ single channel centered images of which $985$ were used for training and 110 for validation. The network was trained for $10,000$ epochs.
 The signal is a sinusoidal with a period of $356$ days and thus it performs three complete cycles in the $1095$ images.\\
 The final MMSE and MSSI on the validation set are found to be respectively $4.4 \times 10^{-8}$ and $0.999938$ and thus the images are reconstructed with a precision good enough (with respect to the accuracy needed) to recover the signal. Although the image reconstruction capability of the network directly correlates with these losses (as one should expect), we don't find any direct correlation with the signal reconstruction capabilities. Despite the fact that after $1000$ epochs the MSE Loss gradient flattened, for some reasons the latent space began showing an increasing sinusoidal trend in time with an increasing number of epochs. To have a loss function that correlates with the signal reconstruction in the latent space, we would need an architecture that makes use of the time dimension of the images: something not anticipated at the time of the publication of this work. For this reason, the network is encoding only the detection of the signal and not its amplitude. In order to make sure that the periodic trend observed in the latent space was actually coming from a signal injected in the images and not from any other periodic trend (in time) in the images or by chance, the Network was run on a blind set of simulations of which some contained a signal and some didn't. The Network latent space didn't show any periodic trend for all the simulations with no signal injected or, even if a period was recorded, the resulting FAP (see Sec. \ref{Signal Analysis}) would be extremely low.
\begin{figure}
    \centering
    \includegraphics[width=0.7\textwidth]{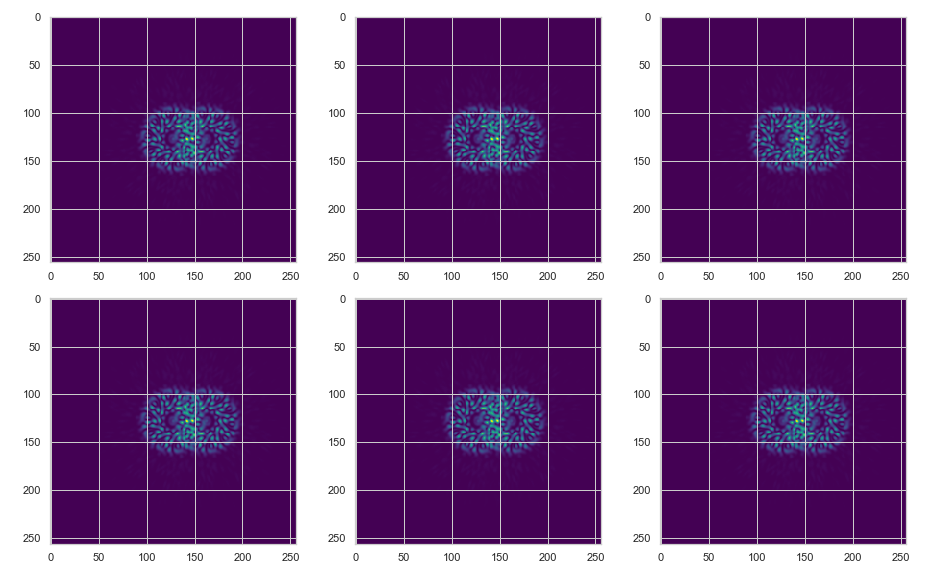}
    \caption{Example of the signal reconstruction after training the network. The first row contains a random subset of simulated Toliman images, the second row shows their respective reconstructions produced by the trained network.}
    \label{fig:reconstruction}
\end{figure}

\begin{figure}[!tbp]
    \centering
        \includegraphics[width=\textwidth]{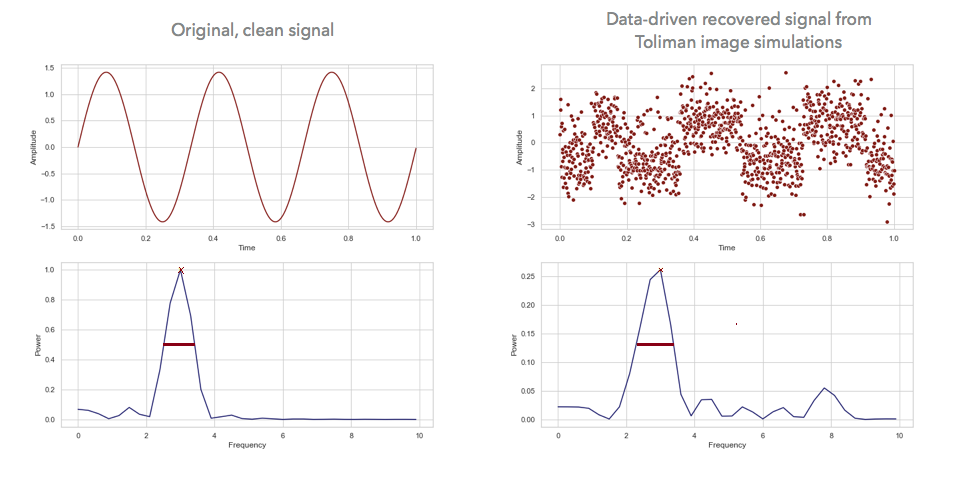} 
    \caption{In the left panels, a perfect signal is represented in the top and the relative Lomb Scargle Periodogram obtained from its analysis is represented in the bottom. In the right, a time series from the atoms obtained with the deep convolutional auto-encoder applied to TOLIMAN simulation with a $10^{-6}$-level astrometric shifts is shown on the top, while its Lomb Scargle Periodogram is represented in the bottom. The power peaks and their relative FWHM are shown in red over the power spectrum.}
    \label{fig:Time Series}
\end{figure}

Table \ref{tab:period_results} shows the result of applying the method of compressing the signal using Deep Convolutional Auto-Encoders (CAE) and afterwards using Lomb Scargle Periodogram to analyse this compressed representation. This table shows that the proposed method is able to capture the signal with very low FAP and reasonable relative error, when compared with the error obtained by a direct analysis of a perfect signal. The Signal and Atom time series, and their relative Lomb Scargle Periodogram, are shown in Fig. \ref{fig:Time Series}. In yellow we highlighted the power peak and the FWHM of the power spectrum around that peak.
\begin{table}
    \centering
    \begin{tabular}{|c|c|c|}
    \hline
      \textbf{Time Series}   &  \textbf{Period} & \textbf{FAP}\\
      \hline
      Signal &  $0.33 \pm 0.05$   & $0$ \\
      \hline
      Atom & $0.33 \pm 0.06$  & $ 7.2 \times 10^{-68}$ \\
      \hline
    \end{tabular}
    \caption{Period found with the Lomb Scargle Periodogram and relative error and FAP.}
    \label{tab:period_results}
\end{table}

\subsection{Discussion of Results}
\label{Discussion}
Sec. \ref{Experiment}  describes both the Network reconstruction capabilities and the analysis on the atom time series to find its periodicity. We have shown that the found periodicity is compatible with the injected astrometric signal period and thus that the architecture is able to recover the signal directly from the TOLIMAN simulation images. One of the main current issues is the lack of correlation between the Network reconstruction of the TOLIMAN images and the presence of a periodic trend in the atom time series. Due to the fact that the Network is only training with spatial information and that the used loss (MSE) only takes into consideration the ability to reconstruct the input images, in reality there is no any encoded reason why the latent space should present a sinusoidal trend with time. The only thing that the latent space should be encoding is "how to reconstruct the images" and nothing else. That being said, given our knowledge of the sinusoidal nature of the astrometric signal and being the astrometric shift the only element changing through the images, we don't see any reason why the latent space should not present a sinusoidal trend with time, regardless the fact that we did not apply any constrain (on the architectural level) to force it. As seen in Fig. \ref{fig:Time Series}, in fact, given enough epochs, the time series actually shows a sinusoidal behaviour.\\
A required step forward in this work is to produce simulations with increasing noise realism and complexity, in order to evaluate if Deep Learning can still be used to recover the astrometric signal. It must be expected that this simple approach would fail to recover the signal if spatial transformations invariance is achieved on an architectural level.

\section{Conclusions}
\label{Conclusions}
In this work we shown how Deep Learning, in particular deep convolutional autoencoders (CAE) can be used to extract, in a completely unsupervised way, periodic astrometric signals with amplitudes of the order of $10^{-6}$ with respect to the size of a pixel. This is the magnitude of the signals that would be produced by an earth-like planet at the habitable zone of a star in the Alpha Centauri binary system (see Sec. \ref{Astrometry}).\\
We presented a detailed explanation of the adopted network architecture (see Sec. \ref{Architecture}) and of the simulations used, which were created using FFT techniques (see Sec. \ref{Simulations}). Although the present simulations do not yet contain certain realistic systematic noise components, such as telescope jitter, rotations and aberrations, they pose a significant challenge to classical unsupervised techniques, due to the small amplitude of the signal with respect to the pixel size. 
We have shown that, from the obtained CAE latent space, we can obtain a time-trend that can be analysed for periodicity, using any time-domain signal extraction technique. Here we used a standard Lomb Scargle technique (see Sec. \ref{Signal Analysis}), and were able to find a period consistent with that of the injected signal (see Sec. \ref{Experiment}).

Finally, we note that in this work we only explored a fully unsupervised method for the compression, although semi-supervised and hybrid methods can be a natural extension, by considering that we may constrain the problem's dimensionality -- for instance, a first order approximation of the shape of the PSF. A further step will be the generation of increasingly realistic systematic noise contributions, to design network architectures that can handle them and still allow for detection of the planetary signal. This work opens an interesting path that we believe should be further studied, towards the extraction of periodic signals of binary systems at the milliarcsecond level, directly from times series of satellite imaging data.

\begin{acknowledgement} This work was partially produced during the $\rm 2^{nd}$ COIN-Focus: Toliman Event\footnote{\url{https://cosmostatistics-initiative.org/focus/toliman1/}} (COIN-Focus \#2) held in Rome, Italy, in November 2019. The COIN-Focus: Toliman participants acknowledge the fundamental support of the Breakthrough Initiatives. The Breakthrough Watch initiative and committee (notably Olivier Guyon, Pete Klupar \& Pete Worden) have supported and framed the problem. We also acknowledge input and ideas from people in the wider TOLIMAN collaboration including Ben Pope, Barnaby Norris, Bryn Jeffries, Anthony Horton and others. MB acknowledges financial contributions from the agreement \textit{ASI/INAF 2018-23-HH.0, Euclid ESA mission - Phase D} and the \textit{INAF PRIN-SKA 2017 program 1.05.01.88.04}. EEOI acknowledges financial support from CNRS 2017 MOMENTUM grant under project \textit{Active Learning for Large Scale Sky Surveys}. AKM acknowledges the support from the Portuguese Funda\c c\~ao para a Ci\^encia e a Tecnologia (FCT) through grants SFRH/BPD/74697/2010, PTDC/FIS-AST/31546/2017 and from the Portuguese Strategic Programme UID/FIS/00099/2013 for CENTRA.
\end{acknowledgement}

\clearpage
\bibliographystyle{plainnat}
\bibliography{bibliography}

\end{document}